\documentclass{sf2a-conf2015}
\usepackage{graphicx}
\usepackage{hyperref}
\usepackage[]{natbib}  
\usepackage{epstopdf}

\usepackage{amsmath, amssymb, amsfonts, amscd}
\usepackage{mathtools} 
\usepackage{array} 
\usepackage[varg]{txfonts} 
\usepackage{multirow}
\usepackage{mathrsfs}

\def\BibTeX{{\rm B\kern-.05em{\sc i\kern-.025em b}\kern-.08em
    T\kern-.1667em\lower.7ex\hbox{E}\kern-.125emX}}
\bibpunct{(}{)}{;}{a}{}{,}  

\begin{document}

\TitreGlobal{SF2A 2015}


\title{Scaling laws to quantify tidal dissipation in star-planet systems}

\runningtitle{Scaling laws to quantify tidal dissipation in star-planet systems}

\author{P. Auclair-Desrotour$^{1,2}$}
\author{S. Mathis$^{2,3}$}
\author{C. Le Poncin-Lafitte}

\address{IMCCE, Observatoire de Paris, UMR 8028 du CNRS, UPMC, 77 Av. Denfert-Rochereau, 75014 Paris, France}
\address{Laboratoire AIM Paris-Saclay, CEA/DSM - CNRS - Universit\'e Paris Diderot, IRFU/SAp Centre de Saclay, F-91191 Gif-sur-Yvette Cedex, France}
\address{LESIA, Observatoire de Paris, CNRS UMR 8109, UPMC, Univ. Paris-Diderot, 5 place Jules Janssen, 92195 Meudon, France}
\address{SYRTE, Observatoire de Paris, UMR 8630 du CNRS, UPMC, 77 Av. Denfert-Rochereau, 75014 Paris, France}

\setcounter{page}{237}


\maketitle


\begin{abstract}
Planetary systems evolve over secular time scales. One of the key mechanisms that drive this evolution is tidal dissipation. Submitted to tides, stellar and planetary fluid layers do not behave like rocky ones. Indeed, they are the place of resonant gravito-inertial waves. Therefore, tidal dissipation in fluid bodies strongly depends on the excitation frequency while this dependence is smooth in solid ones. Thus, the impact of the internal structure of celestial bodies must be taken into account when studying tidal dynamics. The purpose of this work is to present a local model of tidal gravito-inertial waves allowing us to quantify analytically the internal dissipation due to viscous friction and thermal diffusion, and to study the properties of the resonant frequency spectrum of the dissipated energy. We derive from this model scaling laws characterizing tidal dissipation as a function of fluid parameters (rotation, stratification, diffusivities) and discuss them in the context of star-planet systems.
\end{abstract}

\begin{keywords}
hydrodynamics, waves, turbulence, planet-star interactions, planets and satellites: dynamical evolution and stability
\end{keywords}


\section{Introduction}
Planetary fluid layers and stars are affected by tidal perturbations resulting from mutual gravitational and thermal interactions between bodies. These perturbations generate velocity fields which are at the origin of internal tidal dissipation because of the friction/diffusion applied on them. Over long timescales, the energy dissipated in a planetary system impacts the orbital dynamics of this later \citep[][]{EL2007,ADLPM2014}. The architecture of the system thus evolves. At the same time, the rotation of its components and the orientation of their spin is modified while they are submitted to an internal heating. However, solids and fluids are not affected by tides in the same way. While the solid planetary tidal response takes the form of a delayed visco-elastic elongation, internal and external fluid shells such as liquid cores and atmospheres behave as waveguides having their own resonant frequency ranges \citep[][]{OL2004,OL2007,GS2005}. Because of its great complexity, tidal dissipation resulting from this behaviour has been studied in numerous theoretical works, especially for stellar interiors and gaseous giant fluid envelopes, over the past decades \citep[see e.g.][]{Zahn1966a,Zahn1966b,Zahn1966c,Zahn1975,Zahn1977,Zahn1989a,OL2004,Wu2005,OL2007,RMZ2012,Cebron2012,Cebron2013}, which highlighted the crucial role played by the internal structure of bodies and their dynamical properties (rotation, stratification, diffusivities). It is therefore very important to understand the physical mechanisms responsible for tidal dissipation in fluid layers. \\
  
\noindent Tidal waves that can propagate in these layers belong to well-identified families:
\begin{itemize}
\item inertial waves due to the rotation of the body and which have the Coriolis acceleration as restoring force,
\item gravity waves due to the stable stratification of the layers and driven by the Archimedean force,
\item Alfv\'en waves due to magnetic field (if the fluid is magnetized) and driven by magnetic forces.
\end{itemize}

\noindent As demonstrated by \cite{OL2004} for inertial waves, the amplitude of tidal dissipation strongly depends on the tidal frequency contrary to the case of solids. It is also obviously linked to internal properties of the layer such as its turbulent viscosity, thermal diffusivity, rotation and stratification. Indeed, several dissipative mechanisms are involved. The most important of them are viscous friction in turbulent convective zones, thermal diffusion in radiative zones, and Ohmic diffusion in the case of magnetized fluids. In this work, we ignore magnetic effects and focus on gravito-inertial waves damped through viscosity and thermal diffusion. Hence, we give an overview of the analytical results established in \cite{ADMLP2015}. We refer the reader to this paper for more details. Generalizing the approach described by \cite{OL2004}, given in Appendix A of their paper, we consider an idealized local section of a fluid layer submitted to an academic tidal forcing with periodic boundary conditions. This model allows us to compute analytic expressions of energies dissipated by viscous friction and thermal diffusion. Then, we use these results to identify the control parameters of the system, to determine the possible asymptotic regimes of the tidal response and to give simple scaling laws characterizing a dissipation spectrum. Hence, in Sect. 2, we present the local model. We summarize the obtained results in Sect. 3 and give our conclusions in Sect. 4. 

\begin{figure}[b]
 \centering
 \includegraphics[width=0.37\textwidth,clip]{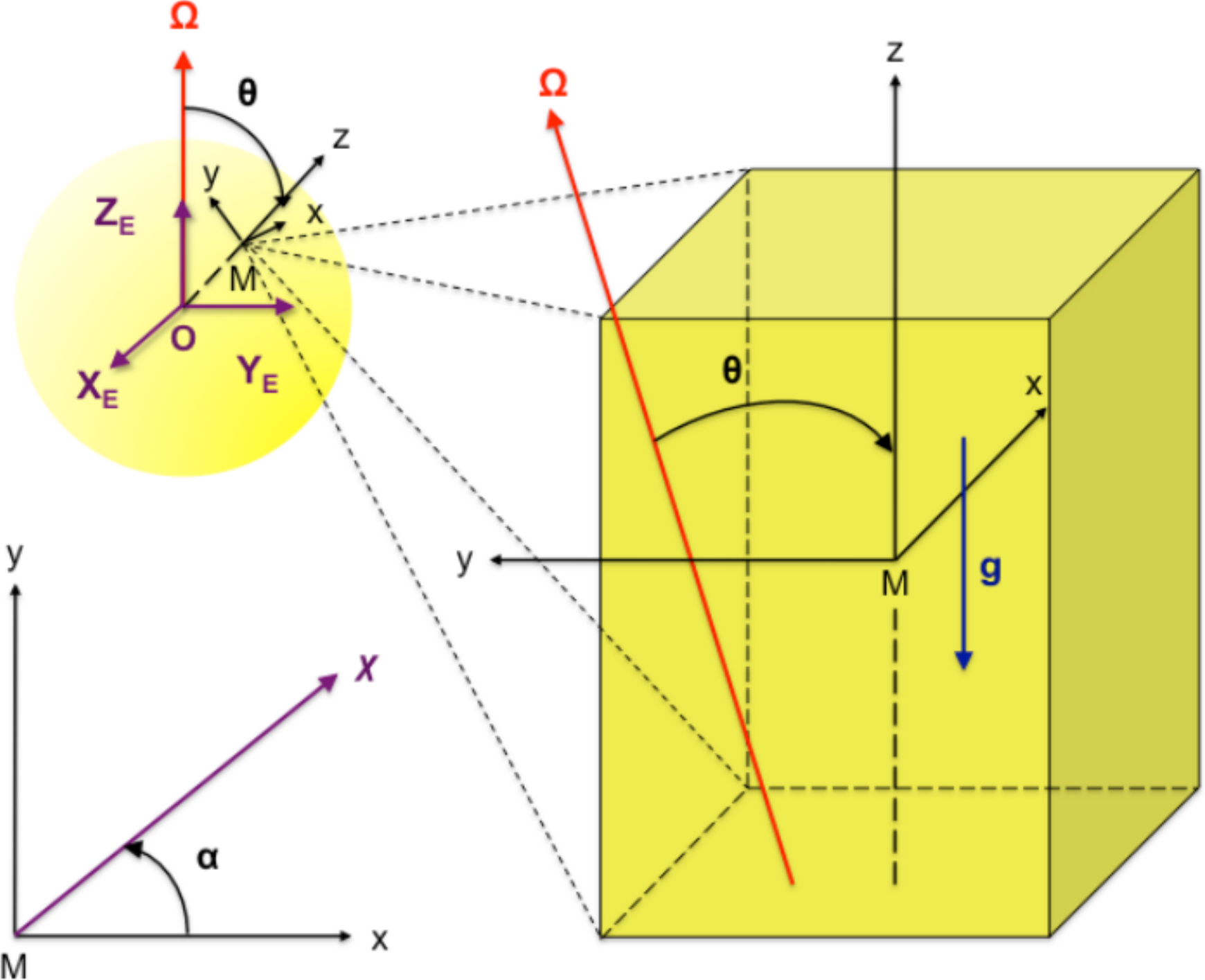}%
 \hspace*{1cm} 
 \includegraphics[width=0.47\textwidth,clip]{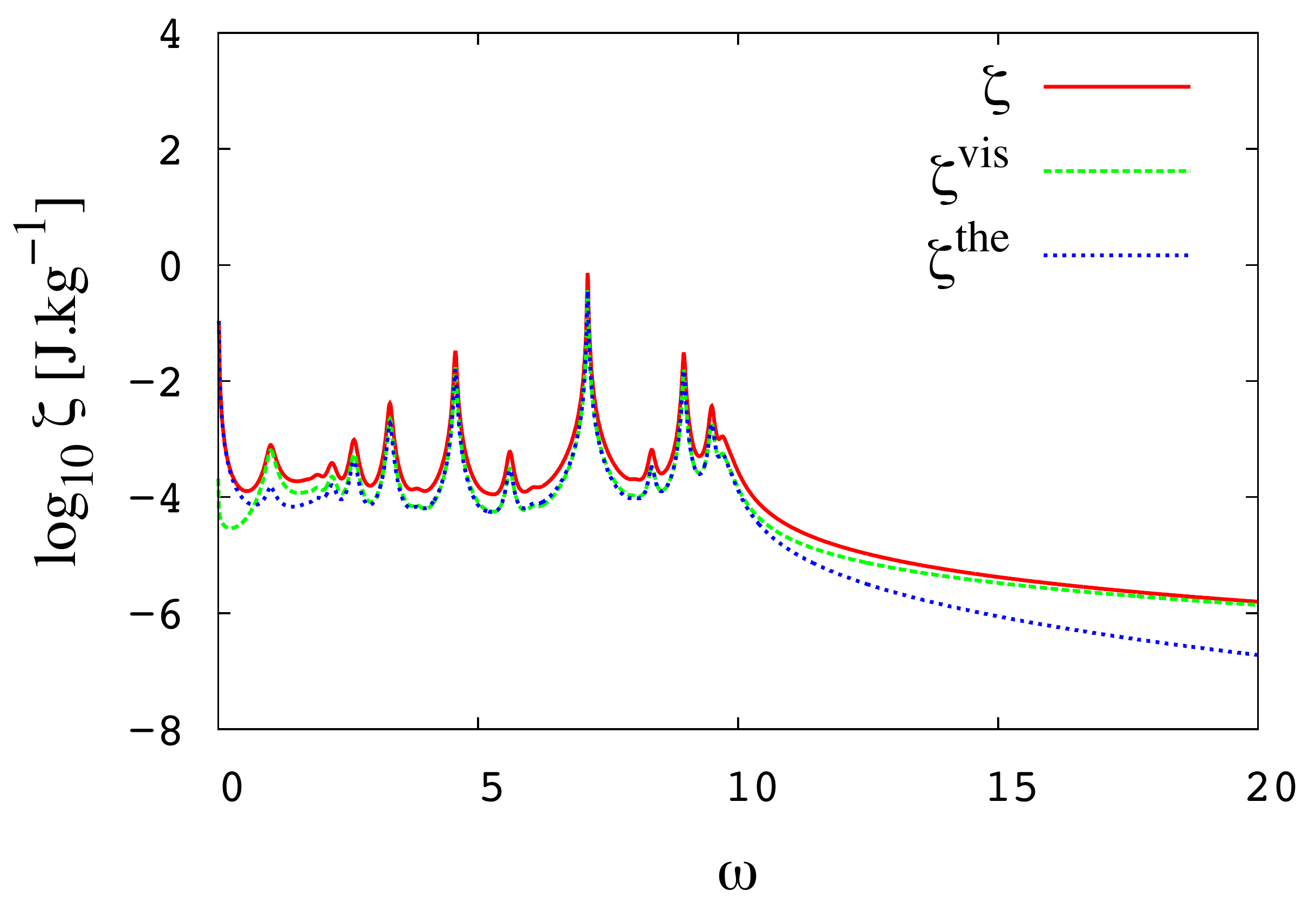}      
  \caption{{\bf Left:} Local Cartesian model, frame, and coordinates. {\bf Right:} Energy dissipated ($\zeta$) and its viscous and thermal components, $ \zeta^{\rm visc} $ and $ \zeta^{\rm therm} $ respectively, as functions of the reduced tidal frequency ($ \omega $) for $ \theta = 0 $, $ A = 10^2 $, $ E = 10^{-4} $ and $ K = 10^{-2} $, which gives $ P_r = 10^{-2} $ (see Sect. 2 for the definition of these quantities). }
  \label{auclair-desrotour:fig1}
\end{figure}
  
\section{Physical set-up}

\subsection{Local model}

Our local model is a Cartesian fluid box of side length $ L $ centered on a point $ M $ of a planetary fluid layer, or star (see Fig.~\ref{auclair-desrotour:fig1}). Let be $ \mathscr{R}_O~:~\left\{O, \textbf{X}_{\rm E} , \textbf{Y}_{\rm E} , \textbf{Z}_{\rm E} \right\} $ the reference frame rotating with the body at the spin frequency $ \Omega $ with respect to $ \textbf{Z}_{\rm E} $.The spin vector $ \boldsymbol{\Omega} $ is thus given by $ \boldsymbol{\Omega}  = \Omega  \textbf{Z}_{\rm E} $. The point $ M $ is defined by the spherical coordinates $ \left( r , \theta , \varphi \right) $ and the corresponding spherical basis is denoted $ \left( \textbf{e}_{r} , \textbf{e}_{\theta} , \textbf{e}_{\varphi} \right) $. We also define the local Cartesian coordinates $ \textbf{x} = \left( x , y , z \right) $ and reference frame $ \mathscr{R}~:~\left\{ M , \textbf{e}_{x} , \textbf{e}_{y} , \textbf{e}_{z}  \right\} $ associated with the fluid box, which is such that $  \textbf{e}_{z} =  \textbf{e}_{r} $, $  \textbf{e}_{x} =  \textbf{e}_{\varphi} $ and $  \textbf{e}_{y} = -  \textbf{e}_{\theta} $. In this frame, the local gravity acceleration, assumed to be constant, is aligned with the vertical direction, i.e. $ \textbf{g} = - g  \textbf{e}_{z} $, and the spin vector is decomposed as follows: $ \boldsymbol{\Omega} = \Omega \left( \cos \theta \textbf{e}_{z} + \sin \theta \textbf{e}_{y} \right) $, where $ \theta $ is the colatitude. The fluid is Newtonian and locally homogeneous, of kinematic viscosity $ \nu $ and thermal diffusivity $ \kappa $. To complete the set of parameters, we introduce the Brunt-V\"ais\"al\"a frequency $ N $ given by

\begin{equation}
N^2 = - g \left[ \dfrac{d \log \rho}{dz}  - \frac{1}{\gamma}  \dfrac{d \log P}{ dz }    \right],
\end{equation}

\noindent where $ \gamma = \left( \partial \ln P / \partial \ln \rho \right)_S $ is the adiabatic exponent ($S$ being the specific macroscopic entropy), and $ P $ and $ \rho $ are the radial distributions of pressure and density of the background, respectively. These distributions are assumed to be rather smooth to consider $ P $ and $ \rho $ constant in the box. The regions studied are stably stratified ($ N^2 > 0 $) or convective ($ N^2 \approx 0 $ or $ N^2 < 0 $). At the end, we suppose that the fluid is in solid rotation with the whole body. 

\subsection{Analytic expressions of dissipated energies}
 
 The fluid is perturbed by a tidal force $ \textbf{F} =\left(F_x ,F_y,F_z\right)$, periodic in time (denoted $ t $) and space, at the frequency $ \chi $. Its tidal response takes the form of local variations of pressure $ p' $, density $ \rho' $, velocity field $ \textbf{u} = \left( u , v , w \right) $ and buoyancy $ \textbf{B} $, which is defined as follows:
 
 \begin{equation}
\textbf{B}=B \textbf{e}_{z}=-{g}\frac{\rho^{'}\left( \textbf{x},t\right)}{\rho}{ \textbf{e}}_{z}\,.
\label{buoyancy}
\end{equation}
  
\noindent Introducing the dimensionless time and space coordinates, tidal frequency, normalized buoyancy, and force per unit mass

\begin{equation}
\begin{array}{c c c c c c c}
   T = 2 \Omega t, & X = \displaystyle \frac{x}{L},& Y=\displaystyle \frac{y}{L}, & Z = \displaystyle \frac{z}{L}, & \omega = \displaystyle \frac{\chi}{2 \Omega}, & \textbf{b} = \displaystyle \frac{\textbf{B}}{2 \Omega}, & \textbf{f} = \displaystyle \frac{ \textbf{F}}{2 \Omega},  \\
\end{array}
\end{equation}

\noindent and using the Navier-Stokes, continuity and heat transport equations, we compute a solution of the tidally forced waves and perturbation, denoted $ s = \left\{ p' , \rho' , \textbf{u}, \textbf{b}, \textbf{f} \right\} $, of the form $ s = \Re \left[ \displaystyle \sum s_{mn} e^{i 2 \pi \left( m X + n Z  \right) } e^{-i \omega T} \right] $, where $ \Re $ stands for the real part of a complex number. In this expression, $ m $ and $ n $ are the longitudinal and vertical degrees of Fourier modes and $ s_{mn} $ the associated coefficient. At the end, the expressions of the energies dissipated per mass unit over a rotation period by viscous friction and thermal diffusion are obtained:

\begin{equation}
\begin{array}{cc}
\zeta^{\rm visc} = 2 \pi E \sum_{ (m,n) \in \mathbb{Z^*}^2 } \left( m^2 + n^2 \right) \left( \left| u_{mn}^2 \right| + \left| v_{mn}^2 \right| + \left| w_{mn}^2 \right| \right), & \zeta^{\rm therm} = 2 \pi K A^{-2}  \sum_{ (m,n) \in \mathbb{Z^*}^2 } \left( m^2 + n^2 \right) \left| b_{mn} \right|^2.
\end{array}
\label{energy}
\end{equation}

\noindent In these expressions, $ A $, $ E $ (the Ekman number) and $ K $ are the control parameters of the system, given by
 
\begin{equation}
\begin{array}{cccc}
A = \left( \displaystyle \frac{N}{2 \Omega}  \right)^2,  &  E = \displaystyle \frac{2 \pi^2 \nu}{ \Omega L^2}, & \mbox{and}  & K = \displaystyle \frac{2 \pi^2 \kappa }{\Omega L^2 }.
\end{array}
\label{control_param}
\end{equation}

\begin{figure}[b]
 \centering
 \includegraphics[width=0.7\textwidth,clip]{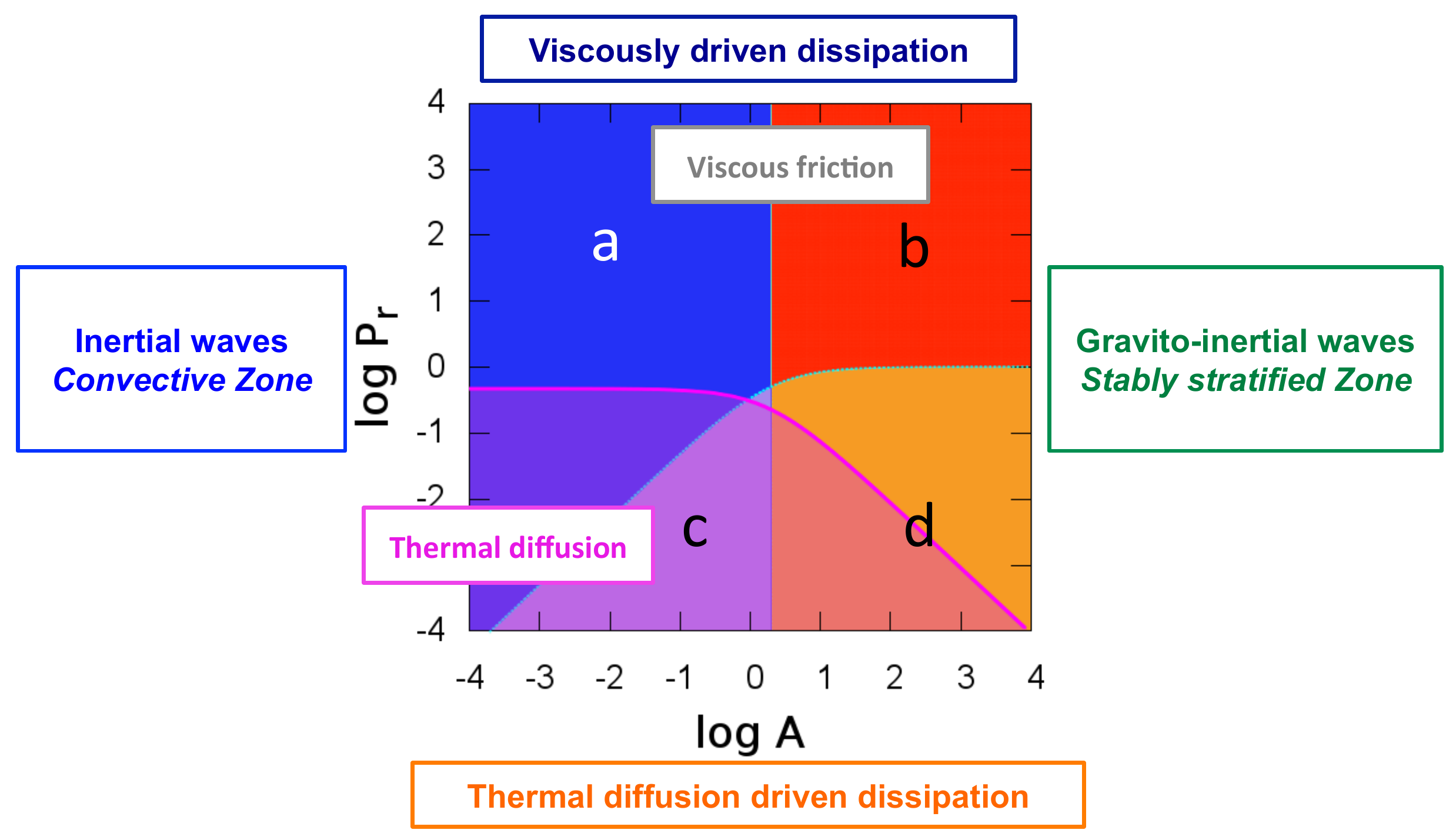}      
  \caption{Map of the asymptotic behaviours of the tidal response. The horizontal (vertical) axis measures the parameter $ A $ ($ P_r = \nu / \kappa $) in logarithmic scales. Regions on the left correspond to inertial waves (\textbf{a} and \textbf{c}), and to gravito-inertial waves on the right (\textbf{b} and \textbf{d}). The fluid viscosity (thermal diffusivity) drives the behaviour of the fluid in regions \textbf{a} and \textbf{b} (\textbf{c} and \textbf{d}). The pink (grey) zone corresponds to the regime of parameters where $ \zeta^{\rm therm} $ ($ \zeta^{\rm visc} $) predominates in tidal dissipation.   }
  \label{auclair-desrotour:fig2}
\end{figure}

\section{Asymptotic regimes and scaling laws}

\begin{figure}[ht!]
 \centering
 \includegraphics[width=0.8\textwidth,clip]{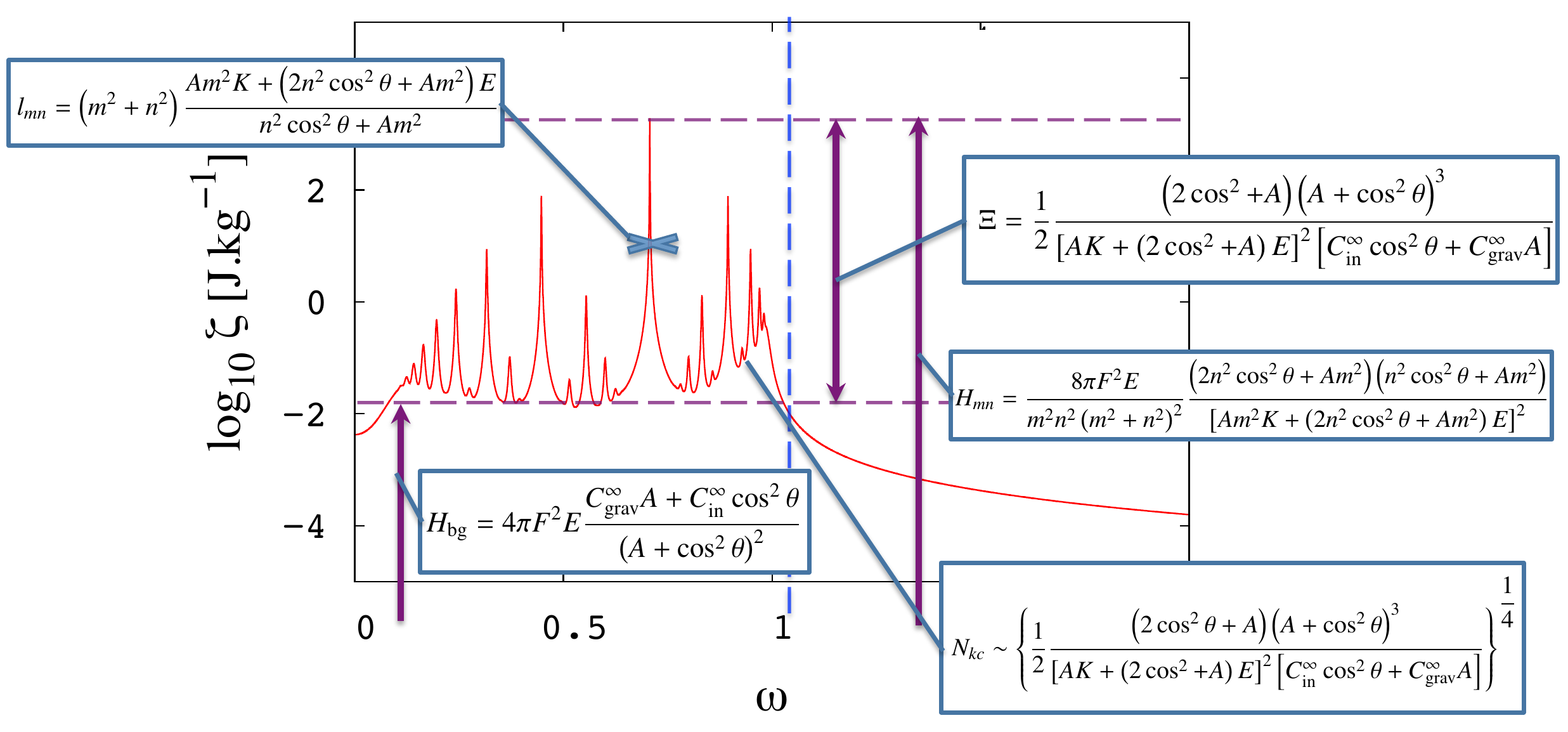}      
  \caption{Energy dissipated by viscous friction as a function of the reduced tidal frequency ($ \omega $) and the formulae giving the properties of the spectrum as functions of the colatitude and control parameters of the box ($ A $, $ E $ and $ K $).  }
  \label{auclair-desrotour:fig3}
\end{figure}

Using Eq.~(\ref{energy}), it is possible to plot $ \zeta^{\rm visc}  $ and $ \zeta^{\rm therm}  $ as functions of the tidal frequency (e.g. Fig.~\ref{auclair-desrotour:fig1}, right). The dissipation spectrum appears to be highly resonant, and its properties strongly depend on the control parameters identified above. By studying the analytic solution given by the model, we determine the asymptotic regimes of the tidal response (Fig.~\ref{auclair-desrotour:fig2}). Let us recall the Prandtl number of the system, $ P_{\rm r} = \nu / \kappa $. Four different behaviours are identified. Each of them corresponds to a colored region on the map:
\begin{itemize}
  \item[1.] $ A \ll A_{mn} $ and $ P_{r} \gg P_{r;mn} $: inertial waves controlled by viscous diffusion (blue);
  \item[2.] $ A \gg A_{mn} $ and $ P_{r} \gg P_{r;mn} $: gravity waves controlled by viscous diffusion (red);
  \item[3.] $ A \ll A_{mn} $ and $ P_{r} \ll P_{r;mn} $: inertial waves controlled by thermal diffusion (purple); and
  \item[4.] $ A \gg A_{mn} $ and $ P_{r} \ll P_{r;mn} $: gravity waves controlled by thermal diffusion (orange),
\end{itemize}

\noindent where $ A_{mn} $ and $ P_{r;mn} $ are the vertical and horizontal transition parameters associated with the mode $ \left( m , n \right) $. Besides, we may identify the regions where the fluid response is mainly damped by thermal diffusion (pink) or by viscous friction (grey). The transition, materialized by the pink line, corresponds to $ P_r = P_r^{\rm diss} $. The model allows us to compute, for all regimes, analytical formulae quantifying properties of the dissipation spectrum such as the number $ N_{\rm kc} $, positions $ \omega_{mn} $, width $ l_{mn} $ and height $ H_{mn} $ of the resonant peaks, the height of the non-resonant background $ H_{\rm bg} $, which corresponds to the equilibrium tide, and the sharpness ratio $ \Xi = H_{11} / H_{\rm bg} $. Some of these formulae are given in Fig.~\ref{auclair-desrotour:fig3}. We finally deduce from these analytic solutions the scaling laws characterizing the dissipation regimes of Fig.~\ref{auclair-desrotour:fig2}, summarized in Table \ref{scaling_laws}.
 

\begin{table*}[htb]
\centering
\begin{tabular}{ | c  |  c    l l  |  c    l l  |}
  \hline
  \hline
    \textsc{Domain} & \multicolumn{3}{c|}{$ A \ll A_{11} $}  & \multicolumn{3}{c|}{$ A \gg A_{11} $}   \\ \hline
     \vspace{0.01mm} & & & & & & \\
    \multirow{5}{*}{$ P_r \gg P_{r;11}^{\rm reg} $} & & $ l_{mn} \propto E $ & $ \omega_{mn} \propto \displaystyle \frac{n  \cos \theta}{\sqrt{m^2 + n^2}} $  &  & $ l_{mn} \propto E $  & $ \omega_{mn} \propto \displaystyle \frac{m \sqrt{A} }{\sqrt{m^2 + n^2}}  $   \\
     \vspace{0.01mm} & & & & & & \\
    & & $ H_{mn} \propto E^{-1}  $  & $ N_{\rm kc} \propto E^{-\frac{1}{2}} $ &  & $ H_{mn} \propto E^{-1} $ &  $ N_{\rm kc} \propto A^{\frac{1}{4}} E^{-\frac{1}{2}} $    \\
     \vspace{0.01mm} & & & & & & \\
    &  & $ H_{\rm bg} \propto E $ & $ \Xi \propto E^{-2}  $ &  & $ H_{\rm bg} \propto A^{-1} E $ & $ \Xi \propto A E^{-2} $    \\
     \vspace{0.01mm} & & & & & & \\
     \hline
     \vspace{0.01mm} & & \multicolumn{1}{| l}{} & & & \multicolumn{1}{| l}{} & \\
   \multirow{12}{*}{$ P_r \ll P_{r;11}^{\rm reg} $} & \multirow{6}{*}{$ P_r \gg P_{r;11} $} & \multicolumn{1}{| l}{$ l_{mn} \propto  E  $} & $ \displaystyle \omega_{mn} \propto \frac{n \cos \theta}{\sqrt{m^2 + n^2}}  $ & \multirow{6}{*}{$ P_r \gg P_{r;11}^{\rm diss} $}  & \multicolumn{1}{| l}{$ l_{mn} \propto E P_r^{-1} $}     & $ \displaystyle \omega_{mn} \propto \frac{m \sqrt{A}}{\sqrt{m^2 + n^2}}    $  \\
     \vspace{0.01mm} & & \multicolumn{1}{| l}{} & & & \multicolumn{1}{| l}{} & \\
    & &  \multicolumn{1}{| l}{$ H_{mn} \propto E^{-1} P_r^{-1} $} & $ N_{\rm kc} \propto E^{-\frac{1}{2}}  $  & & \multicolumn{1}{| l}{$ H_{mn} \propto E^{-1} P_r^2 $} & $ N_{\rm kc} \propto A^{\frac{1}{4}} E^{-\frac{1}{2}} P_r^{\frac{1}{2}}  $   \\
     \vspace{0.01mm} & & \multicolumn{1}{| l}{} & & & \multicolumn{1}{| l}{} & \\
    & &  \multicolumn{1}{| l}{$ H_{\rm bg} \propto E P_r^{-1}  $} & $ \Xi \propto E^{-2} $ & & \multicolumn{1}{| l}{$ H_{\rm bg} \propto A^{-1} E $} & $ \Xi \propto A E^{-2} P_r^{2} $   \\
     \vspace{0.01mm} & & \multicolumn{1}{| l}{} & & & \multicolumn{1}{| l}{} & \\
         \cline{2-7} 
     \vspace{0.01mm} & & \multicolumn{1}{| l}{} & & & \multicolumn{1}{| l}{} & \\
      & \multirow{6}{*}{$ P_r \ll P_{r;11} $} & \multicolumn{1}{| l}{$ l_{mn} \propto A E P_r^{-1} $} & $ \displaystyle \omega_{mn} \propto \frac{n \cos \theta}{\sqrt{m^2 + n^2}}  $ & \multirow{6}{*}{$ P_r \ll P_{r;11}^{\rm diss} $}  & \multicolumn{1}{| l}{$ l_{mn} \propto E P_r^{-1}  $}  & $ \displaystyle \omega_{mn} \propto \frac{m \sqrt{A}}{\sqrt{m^2 + n^2}}   $  \\
     \vspace{0.01mm} & & \multicolumn{1}{| l}{} & & & \multicolumn{1}{| l}{} & \\
    & &  \multicolumn{1}{| l}{$ H_{mn} \propto A^{-2} E^{-1} P_r $} & $ N_{\rm kc} \propto A^{-\frac{1}{2}} E^{-\frac{1}{2}} P_r^{\frac{1}{2}}   $ & & \multicolumn{1}{| l}{$ H_{mn} \propto A^{-1} E^{-1} P_r $} & $ N_{\rm kc} \propto A^{\frac{1}{4}} E^{-\frac{1}{2}} P_r^{\frac{1}{2}}  $   \\
     \vspace{0.01mm} & & \multicolumn{1}{| l}{} & & & \multicolumn{1}{| l}{} & \\
    & & \multicolumn{1}{| l}{$ H_{\rm bg} \propto E P_r^{-1} $} & $ \Xi \propto A^{-2} E  $ & & \multicolumn{1}{| l}{$ H_{\rm bg} \propto A^{-2} E P_r^{-1} $} & $ \Xi \propto A E^{-2} P_r^2 $   \\
     \vspace{0.01mm} & & \multicolumn{1}{| l}{} & & & \multicolumn{1}{| l}{} & \\
  \hline
  \hline
\end{tabular}
\caption{\label{scaling_laws} Scaling laws for the properties of the energy dissipated in the different asymptotic regimes. The parameter $ P_{r;11}^{\rm diss} $ indicates the transition zone between a dissipation led by viscous friction and a dissipation led by heat diffusion. The parameter $ A_{11} $ indicates the transition between tidal inertial and gravity waves. The parameter $ P_{r;11}^{\rm reg} $ is defined as $ P_{r;11}^{\rm reg} = {\rm max} \left\{ P_{r;11} , P_{r;11}^{\rm diss} \right\} $.}
 
 \end{table*}


\section{Conclusions}

In this work, we have explored the physics of tidal dissipation in fluid layers by using an analytic local model. This approach allowed us to identify the physical parameters that control the tidal response of a non-magnetized fluid. From the analytic expressions obtained for energies, we determined the possible regimes of tidal dissipation, which may be dominated either by inertial or gravity waves, and controlled either by viscous friction or thermal diffusion. Furthermore, we note that below a given critical Prandtl number, the principal damping mechanism is heat diffusion (on the contrary, tidal dissipation above this Prandtl number is due to viscous friction essentially). At the end, we established the scaling laws quantifying the properties of dissipation frequency spectra as functions of the control parameters of the model for each identified behaviour. This study will be completed in forthcoming works with the case of magnetized fluid layers.

\begin{acknowledgements}
This work was supported by the French Programme National de Plan\'etologie (CNRS/INSU), the CNES-CoRoT grant at CEA-Saclay, the "Axe f\'ed\'erateur Etoile" of Paris Observatory Scientific Council, and the International Space Institute (ISSI; team ENCELADE 2.0).
\end{acknowledgements}

\bibliographystyle{aa}  
\bibliography{auclair-desrotour} 

\begin{thebibliography}{16}
\expandafter\ifx\csname natexlab\endcsname\relax\def\natexlab#1{#1}\fi

\bibitem[{{Auclair-Desrotour} {et~al.}(2014){Auclair-Desrotour}, {Le
  Poncin-Lafitte}, \& {Mathis}}]{ADLPM2014}
{Auclair-Desrotour}, P., {Le Poncin-Lafitte}, C., \& {Mathis}, S. 2014, \aap,
  561, L7

\bibitem[{{Auclair Desrotour} {et~al.}(2015){Auclair Desrotour}, {Mathis}, \&
  {Le Poncin-Lafitte}}]{ADMLP2015}
{Auclair Desrotour}, P., {Mathis}, S., \& {Le Poncin-Lafitte}, C. 2015, \aap,
  581, A118

\bibitem[{{C{\'e}bron} {et~al.}(2013){C{\'e}bron}, {Bars}, {Gal}, {Moutou},
  {Leconte}, \& {Sauret}}]{Cebron2013}
{C{\'e}bron}, D., {Bars}, M.~L., {Gal}, P.~L., {et~al.} 2013, \icarus, 226,
  1642

\bibitem[{{C{\'e}bron} {et~al.}(2012){C{\'e}bron}, {Le Bars}, {Moutou}, \& {Le
  Gal}}]{Cebron2012}
{C{\'e}bron}, D., {Le Bars}, M., {Moutou}, C., \& {Le Gal}, P. 2012, \aap, 539,
  A78

\bibitem[{{Efroimsky} \& {Lainey}(2007)}]{EL2007}
{Efroimsky}, M. \& {Lainey}, V. 2007, Journal of Geophysical Research
  (Planets), 112, 12003

\bibitem[{{Gerkema} \& {Shrira}(2005)}]{GS2005}
{Gerkema}, T. \& {Shrira}, V.~I. 2005, Journal of Fluid Mechanics, 529, 195

\bibitem[{{Ogilvie} \& {Lin}(2004)}]{OL2004}
{Ogilvie}, G.~I. \& {Lin}, D.~N.~C. 2004, \apj, 610, 477

\bibitem[{{Ogilvie} \& {Lin}(2007)}]{OL2007}
{Ogilvie}, G.~I. \& {Lin}, D.~N.~C. 2007, \apj, 661, 1180

\bibitem[{{Remus} {et~al.}(2012){Remus}, {Mathis}, \& {Zahn}}]{RMZ2012}
{Remus}, F., {Mathis}, S., \& {Zahn}, J.-P. 2012, \aap, 544, A132

\bibitem[{{Wu}(2005)}]{Wu2005}
{Wu}, Y. 2005, \apj, 635, 688

\bibitem[{{Zahn}(1966{\natexlab{a}})}]{Zahn1966a}
{Zahn}, J.~P. 1966{\natexlab{a}}, Annales d'Astrophysique, 29, 313

\bibitem[{{Zahn}(1966{\natexlab{b}})}]{Zahn1966b}
{Zahn}, J.~P. 1966{\natexlab{b}}, Annales d'Astrophysique, 29, 489

\bibitem[{{Zahn}(1966{\natexlab{c}})}]{Zahn1966c}
{Zahn}, J.~P. 1966{\natexlab{c}}, Annales d'Astrophysique, 29, 565

\bibitem[{{Zahn}(1975)}]{Zahn1975}
{Zahn}, J.-P. 1975, \aap, 41, 329

\bibitem[{{Zahn}(1977)}]{Zahn1977}
{Zahn}, J.-P. 1977, \aap, 57, 383

\bibitem[{{Zahn}(1989)}]{Zahn1989a}
{Zahn}, J.-P. 1989, \aap, 220, 112

\end{thebibliography}

\end{document}